\documentclass[11pt,preprint]{aastex}

\usepackage{amsmath}
\usepackage{graphicx}
\usepackage{natbib}
\usepackage{hyperref}
\usepackage[usenames,dvipsnamens]{xcolor}
\usepackage[encapsulated]{CJK}
\usepackage{ucs}
\usepackage[utf8x]{inputenc}

\newcommand{\AU}{{\rm AU}}
\newcommand{\R}{{\bf r}}
\newcommand{\V}{{\bf v}}
\newcommand{\x}{{\bf x}}
\newcommand{\s}{{\bf s}}
\newcommand{\den}{{\cal Q}}
\newcommand{\half}{{\frac{1}{2}}}
\newcommand{\Dmin}{{D_{\rm min}}}
\newcommand{\krtext}[1]{\begin{CJK}{UTF8}{mj}#1\end{CJK}}
\begin{document}

\title{Simplified derivation of the collision probability of two objects in independent Keplerian orbits}

\author{Youngmin JeongAhn (\krtext{정안영민})\altaffilmark{1,2} and Renu Malhotra\altaffilmark{2}}
\altaffiltext{1}{Instituto de Astronom\'{i}a, Universidad Nacional Aut\'{o}noma de M\'{e}xico, Apdo. Postal 106, Ensenada, B.C. 22860 M\'{e}xico.}
\altaffiltext{2}{Lunar and Planetary Laboratory, The University of Arizona, Tucson, AZ 85721, USA.}
\email{jeongahn@astro.unam.mx , renu@lpl.arizona.edu}

\begin{abstract}

Many topics in planetary studies demand an estimate of the collision probability of two objects moving on nearly Keplerian orbits.  In the classic works of \citet{Opik:1951} and \citet{Wetherill:1967}, the collision probability was derived by linearizing the motion near the collision points, and there is now a vast literature using their method.
We present here a simpler and more physically motivated derivation for non-tangential collisions in Keplerian orbits, as well as for tangential collisions that were not previously considered.  Our formulas have the added advantage of being manifestly symmetric in the parameters of the two colliding bodies. In common with the {\"O}pik-Wetherill treatments, we linearize the motion of the bodies in the vicinity of the point of orbit intersection (or near the points of minimum distance between the two orbits) and assume a uniform distribution of impact parameter within the collision radius.  We point out that the linear approximation leads to singular results for the case of tangential encounters.  We regularize this singularity by use of a parabolic approximation of the motion in the vicinity of a tangential encounter.
\end{abstract}

\section{Introduction}
An accurate estimate of the intrinsic collision probability between two objects moving on independent Keplerian orbits is essential in many topics in planetary system studies: the impact flux of interplanetary dust particles and minor planets on the Earth and other planets \citep{Opik:1951,Moses:1999,Ivanov:2001}, evolution of the orbits of a swarm of planetesimals \citep{Greenberg:1982}, planet formation \citep{Wetherill:1990}, dynamical lifetimes of small bodies \citep{Dones:1999}, 
collisional evolution of asteroids \citep{Bottke:2005}, the impact hazard of near-Earth asteroids \citep{Harris:2015}, and collisions amongst artificial Earth-orbiting satellites \citep{Liou:2006}.

In such problems, one usually wishes to quantify the probability of collision within some volume of space which is small compared to the uncertainties of the orbital parameters of the objects that pass through that volume; the objects are assumed to be moving on independent Keplerian orbits about a central body.
\citet{Opik:1951} derived an equation for this collision probability for the case when one of the objects is assumed to be in a circular orbit. \citet{Wetherill:1967} generalized this solution to  two eccentric orbits. {\"O}pik's and Wetherill's approaches have two steps in the calculation of collision probability. First, the collision probability for two intersecting Keplerian orbits is calculated. We call this probability $P$, which is a function of the collision radius and the orbital elements defining the shape and the orientation of the two orbits. 
The two orbits are assumed to be fixed in space, and the mean anomalies are assumed to be independent (i.e., there is no mean motion resonance between the two bodies).  Then, over a long period of time, the pair of objects has a well-defined probability of impact near the location where the two orbits intersect  or where the distance between the two orbits is small enough for collision to be possible. (The collision radius is inflated from the sum of the physical radii of the bodies, to account for the gravitational focusing effect from their mutual interaction.)
Secondly, for ensembles of bodies, the equation for $P$ works as a back-bone to calculate the average collision probability marginalized over all values of the mutual argument of pericenter, $\omega$.  For most values of $\omega$, the minimum distance between two orbits is much larger than the distance that allows collision. Thus, the specific ranges of $\omega$ that allow the collision condition is calculated and $P$ over those intervals is integrated over the entire range.

In most previous works in the context of collision rates of asteroids, it has been assumed that apsidal and nodal precession rates are uniform so that $\omega$ is uniformly distributed over its range $(0,2\pi)$ \citep{Wetherill:1967,Greenberg:1982,Bottke:1993}. In two recent studies \citep{Vokrouhlicky:2012,Pokorny:2013}, the assumption of uniform precession is discarded and the secular evolution along the Kozai-Lidov cycle is adopted to integrate $P$ for high inclination orbits. \citet{Rickman:2014} showed that a Monte-Carlo method can be used to integrate $P$ over the precession cycle. Although all the cases described in \citet{Rickman:2014} have uniformly distributed angular parameters, their method can be straightforwardly extended to applications with non-uniformly distributed angular parameters.  \citet{JeongAhn:2015} implemented the {\"O}pik-Wetherill method for the case of non-uniform precession by generating a large number of clones which follow the non-uniform angular distributions. The $P$ values of clones are then summed up to yield the total collision probability on a target object without assuming uniform precession or uniform distribution of $\omega$. This method was used to calculate the seasonal variation of the asteroid impact flux on Mars \citep{JeongAhn:2015}.

We stress that collision probability, $P$, should be interpreted only from the statistical standpoint of collisions of a large population of small bodies in nearly Keplerian orbits. It is not appropriate for predicting a specific impact event in the near future, nor for estimating the long-time impact probability, for a specific pair of objects. For the former purpose, we need the information of passage time which is regarded as a random value in the calculation of $P$. For the latter purpose, the premise of fixed Keplerian orbits is invalid as orbits evolve over time. With the steady state condition for the orbital distribution of numerous colliders, however, we can statistically calculate the impact probability for a given object by integrating $P$ over the distribution of colliders.

In the present work, we first present, in Section~\ref{s:nontngcol}, a new and simplified derivation of \citet{Wetherill:1967}'s collision probability, $P$.  We then show that $P$ diverges when the two objects are moving in the same direction, i.e., when the two bodies have a tangential encounter.  Several authors \citep{Greenberg:1982,Vokrouhlicky:2012,Rickman:2014} have pointed out this singularity in \citet{Opik:1951} and \citet{Wetherill:1967}'s approach, but  addressed it only in the averaging of $P$ over the precession cycle. \citet{Greenberg:1988} discussed outcomes of tangential encounters but did not calculate the modification of $P$ in such cases. Thus, the singularity problem in $P$ itself has not been solved. Even though near-tangential encounters are not very common, the singularly high collision probability of even just a few such cases can cause non-negligible errors in estimates of collision rates. We examine this singularity problem carefully, and we derive an improved equation for $P$ for tangential encounters which regularizes this singularity (Section~\ref{s:tangenc}).  We comment on the practical implementation of the formulas for tangential and non-tangential encounters in a general purpose code for collision rates (Section~\ref{s:transition}), and we describe a case study to illustrate the importance  of the correct treatment of tangential encounters (Section~\ref{s:casestudy}). We summarize and conclude in Section~\ref{s:summary}.

\section{Collision Probability for Non-tangential Encounters\label{s:nontngcol}}
\subsection{Derivation for Intersecting Orbits\label{s:drvintsec}}
\begin{figure}
\centering
  \includegraphics[width=300px]{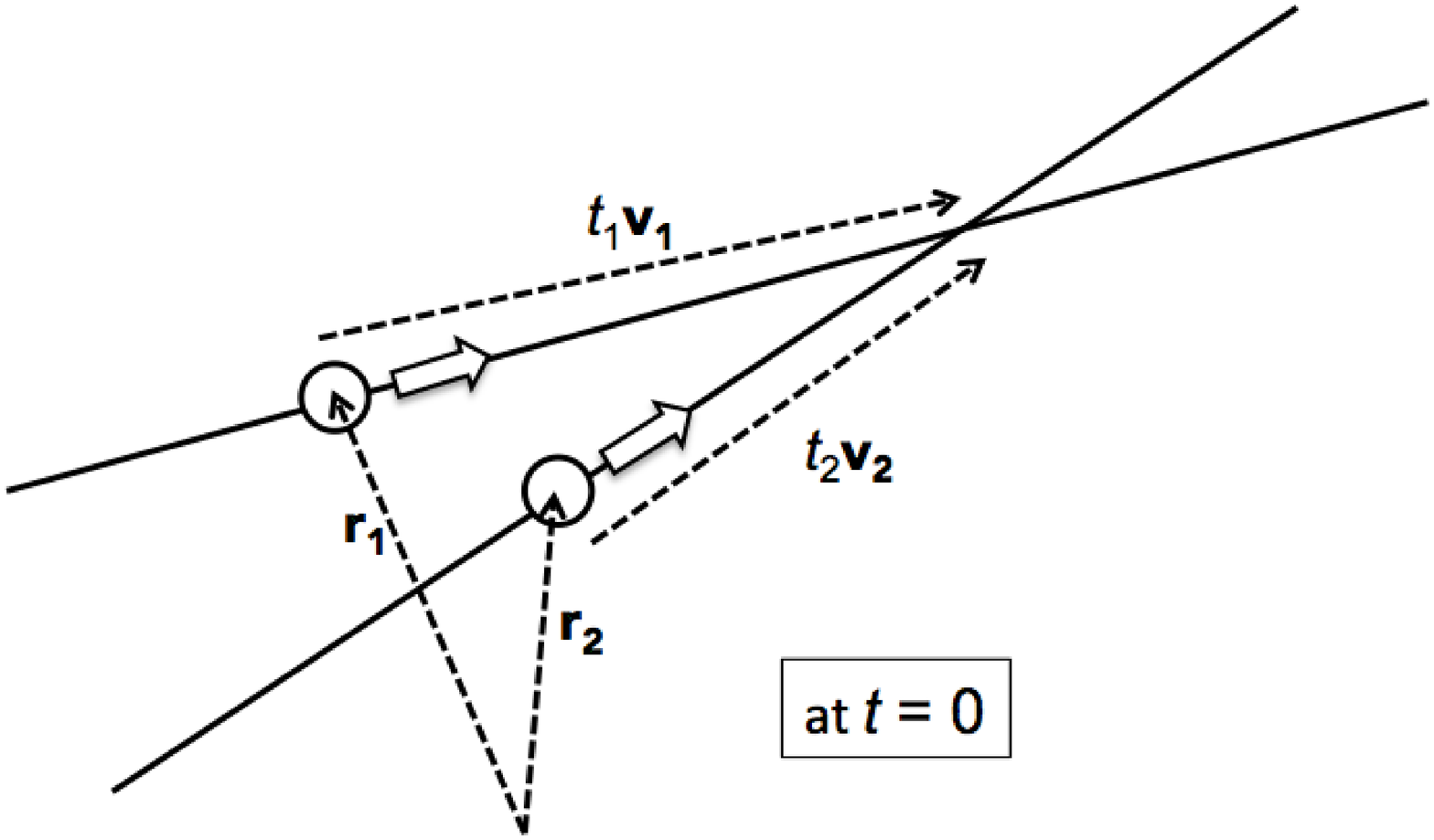}
  \caption{Diagram of two intersecting orbits. Linear trajectories of body 1 and body 2 are intersecting each other but their minimum distance with time is generally not located at the intersection point. }
\label{LNTRJ}
\end{figure}
Consider the motion of two bodies, body 1 and body 2, whose fixed Keplerian orbits are intersecting each other. Their motions are approximated to be linear near the orbit intersection where collision is possible,  as illustrated in Figure~\ref{LNTRJ}. Each body's position can be written as 
\begin{equation}
{\boldsymbol {\rho}}(t) = {\bf r} + t {\bf v}, 
\label{rho}
\end{equation}
where ${\bf r}$ is an arbitrary position vector along the line at time $t=0$ and ${\bf v}$ is the constant velocity in the neighborhood of the orbit intersection point. We can set up a position relation between the two bodies using the point of intersection:
\begin{equation}
{\bf r_1} + t_1 {\bf v_1} = {\bf r_2} + t_2 {\bf v_2},
\label{rtv}
\end{equation}
where subscripts denote the object number (Figure~\ref{LNTRJ}). In this configuration each body passes the orbit intersection point at time $t_1$ and time $t_2$, respectively. 
The encounter velocity is
\begin{equation}
{\bf U}  = {\bf v_1} - {\bf v_2}.
\end{equation}
By taking the cross product with the encounter velocity on both sides of Equation~\ref{rtv} and re-arranging the terms, we get
\begin{equation}
({\bf r_1} - {\bf r_2} ) \times {\bf U} = (t_1 - t_2 ) ({\bf v_1} \times {\bf v_2} ).
\label{rXvrel}
\end{equation}
Any pair of bodies moving with non-parallel constant velocities has a unique time when the distance between the two bodies,  $\left| {\bf r_1} + t {\bf v_1} - {\bf r_2} - t {\bf v_2} \right|$, becomes minimum. By calling this specific time $t=0$, we get the minimum distance $\Dmin = \left| {\bf r_1} - {\bf r_2} \right|$. 

At the minimum distance, the encounter velocity vector is normal to the relative position vector.  This is a trivial consequence of the fact that at the minimum mutual distance,
$$ {d\over dt} |\boldsymbol{\rho}_1(t)-\boldsymbol{\rho}_2(t)|^2 = 0,$$
and the left hand side is proportional to $(\R_1-\R_2)\cdot\bf{U}$.  As this concept is a core part in further derivations, we write it down as a mathematical theorem below for frequent reference.
\newtheorem{theorem}{Theorem}
\begin{theorem}
\label{t:normal}
Two moving points have their local minimum distance when their relative position vector is normal to their encounter velocity vector.
\end{theorem}
This is true for any two moving points on regular curves, except when the relative distance of the two moving points vanishes and the direction of their relative position becomes meaningless.
We mention that Theorem~\ref{t:normal} encapsulates the same concept as in the definition of the so-called ``$b$-plane" adopted by \citet{Greenberg:1988} in the context of \"Opik-Wetherill formulas. The $b$-plane is defined as the plane passing through the target body and being normal to the inbound asymptotic (unperturbed) relative velocity. 
The impact parameter $b$, which is the magnitude of the projection of the inbound asymptote on the $b$-plane, is related to the minimum encounter distance by a scaling factor that accounts for the gravitational interaction between the two bodies.  
In fact, the gravitational interaction between the bodies makes the two bodies approach closer than the distance $b$ (this is the so-called ``gravitational focusing" effect); we will include this effect when we discuss the collision radius later. We note that gravitational focusing also causes a change in the relative velocity of the two bodies, but this is usually insignificant and is neglected in the  calculation of $P$ in the \"Opik-Wetherill approach.  Throughout this paper, we use ``encounter velocity" to refer to the relative velocity which is unperturbed by the mutual gravitational interaction of body 1 and body 2. 

From Theorem~\ref{t:normal} and Equation~\ref{rXvrel}, we can find that the time interval between body 1 and body 2 passing the orbit intersection point, $\Delta t=\left| t_1 - t_2 \right|$, is given by
\begin{equation}
\Delta t = \frac{\Dmin U }{\left| {\bf v_1} \times {\bf v_2} \right|},
\label{Dt1}
\end{equation}
where $U$ is the  scalar  magnitude of ${\bf U}$. 

\begin{figure}
\centering
  \includegraphics[width=300px]{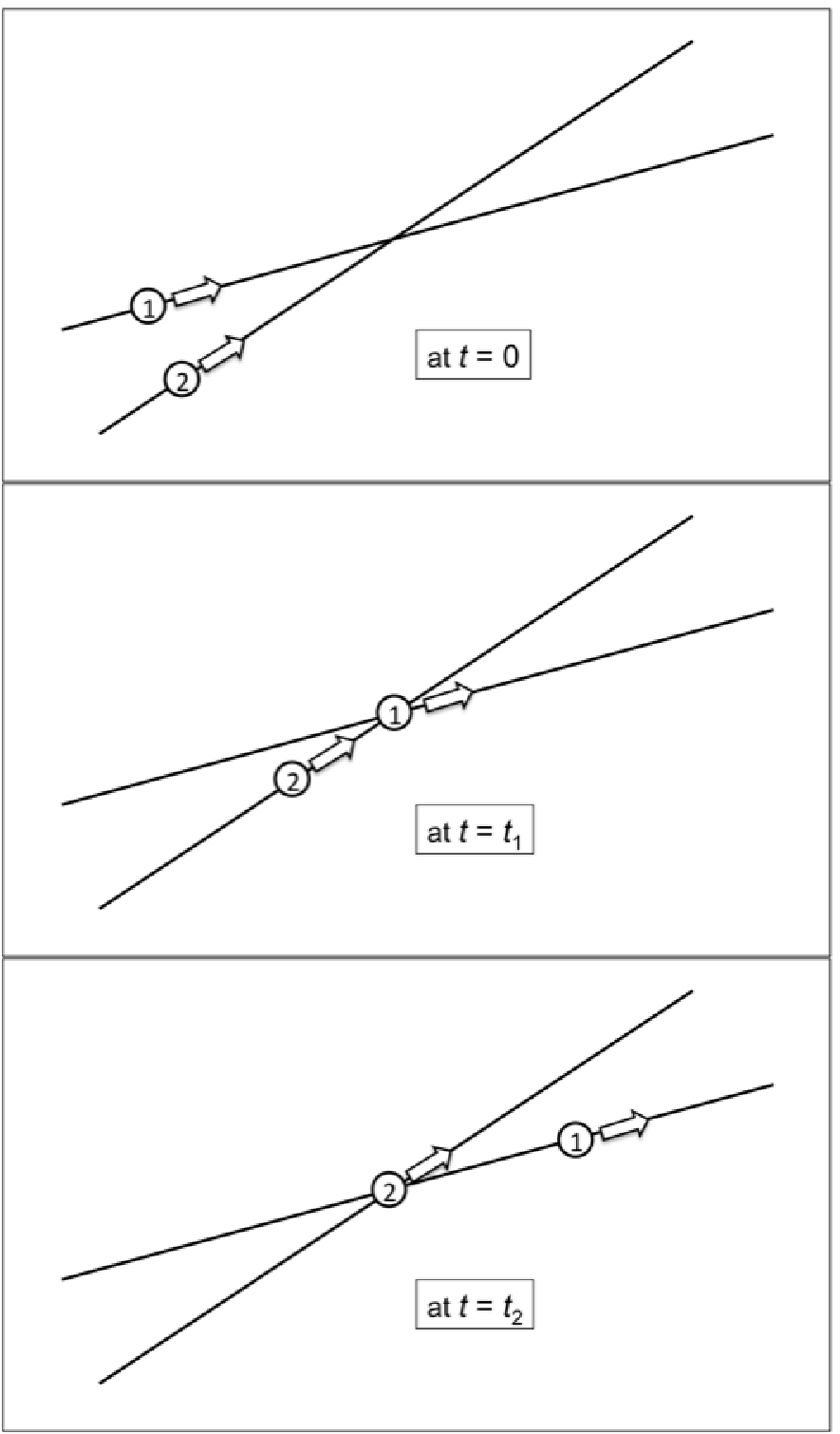}
  \caption{Time slice of two intersecting orbits as Figure~\ref{LNTRJ}. The two bodies have minimum distance, $\Dmin$, at $t=0$.}
\label{BulletT}
\end{figure}

For convenience we choose body 1 as the body passing the intersection point ahead of the other body ($t_1<t_2$), and illustrate this in Figure~\ref{BulletT}. Note that both $t_1$ and $t_2$ can have negative values as one or both bodies may have already passed the orbit intersection point when the two bodies approach each other with the minimum distance, {\it i.e.} when $t=0$. In the middle panel of Figure~\ref{BulletT}, body 2 is $\Delta t$ ahead of the orbit intersection point when body 1 is passing the intersection ($t=t_1$). If body 2 were farther away from the intersection at $t=t_1$, the pair would have larger minimum distance than $\Dmin$. Likewise, if body 2 were located near the orbit intersection point at $t=t_1$, the minimum distance would be smaller than $\Dmin$. Collision is possible if $\Dmin$ is smaller than a specified collision radius $\tau$, {\it i.e.} $\Delta t < \Delta t_{\rm col}$, where $\Delta t_{\rm col}$ is given by
\begin{equation}
\Delta t_{\rm col} = \frac{\tau U}{\left| {\bf v_1} \times {\bf v_2} \right|}.
\label{Dtcol}
\end{equation}
 The collision radius $\tau$ would be the sum of the physical radii of the two bodies, if we neglect their mutual gravity.  If we wish to take account of their mutual gravity, we can multiply with the gravitational focusing factor.
 
Thus, if body 2 passes the orbit intersection point later than body 1, collision would occur if and only if, at time $t=t_1$, body 2 is within $\Delta t_{\rm col}$ of reaching the intersection point. Conversely, if body 2 passes the orbit intersection point ahead of body 1, collision would occur if and only if body 2 already has passed within the time interval $\Delta t_{\rm col}$ before body 1 passes the intersection point. Therefore, whenever body 1 passes the intersection point, collision would occur if body 2 is within the time interval $2 \Delta t_{\rm col}$ of reaching the intersection point.  

Considering that body 1 passes the intersection point once per its orbital revolution, the probability $P_1$ that it collides with body 2 in any single orbital period is
\begin{equation}
P_1 = \frac{2 \Delta t_{\rm col}}{T_2}.
\end{equation}
Thus the collision probability per unit time, $P$, for body 1 to collide with body 2 is $P_1$ divided by the orbital period of body 1, 
\begin{equation}
P = \frac{2 \Delta t_{\rm col}}{T_1 T_2} = \frac{2 \tau U}{\left| {\bf v_1} \times {\bf v_2} \right| T_1 T_2} .
\label{Pmoid0}
\end{equation}
Note that $P$ is the collision probability per unit time when two orbits exactly intersect. The general case of non-intersecting orbits will be considered in the next section.

\begin{figure}
\centering
  \includegraphics[width=300px]{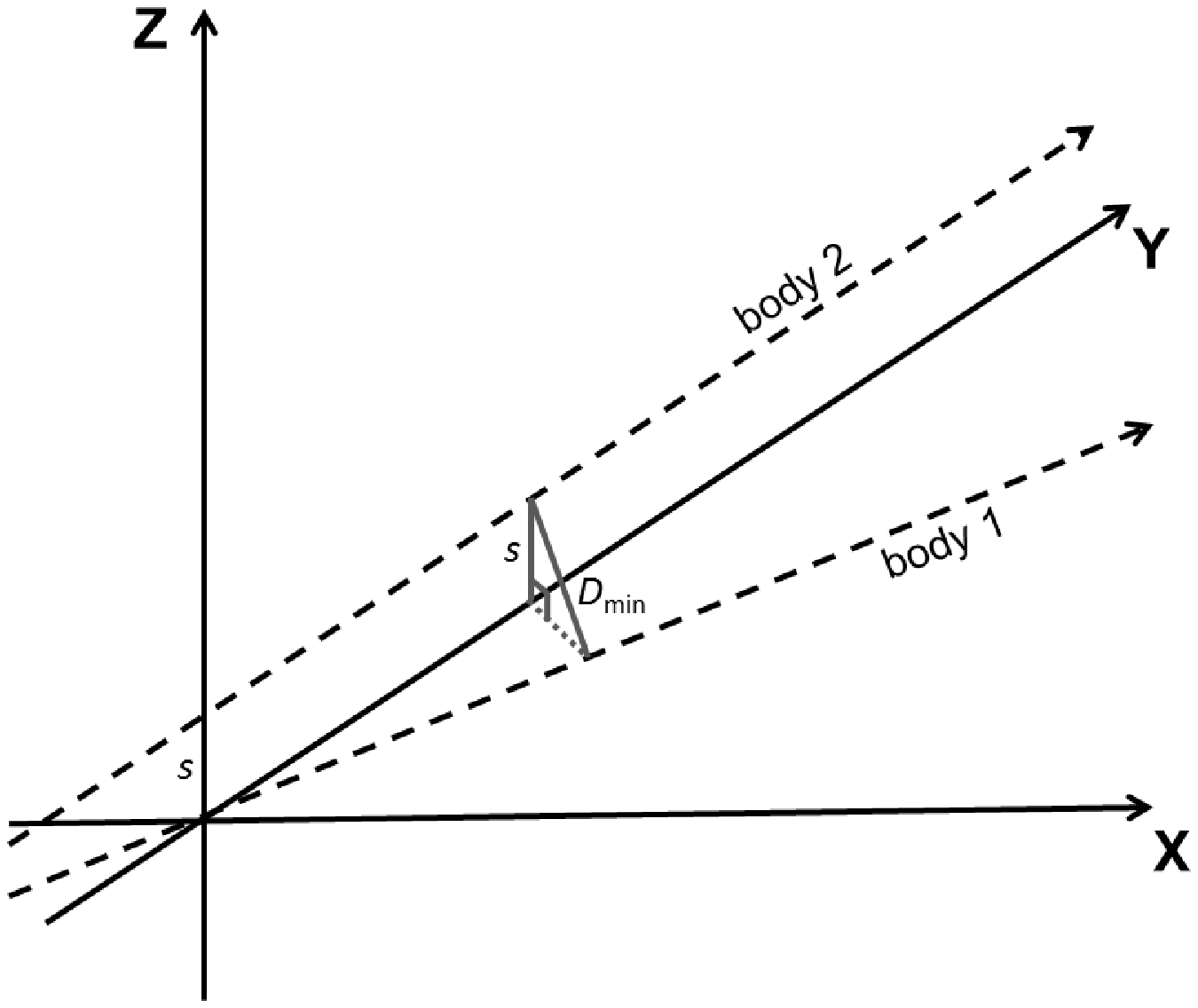}
  \caption{Diagram of colliding bodies, body 1 and body 2, moving in non-intersecting orbits. The Minimum Orbit Intersection Distance (MOID) between two orbits, $s$, is along the Z axis.  The Y direction is chosen to be parallel to the direction of motion of body 2. The XY plane is the plane of the orbit of body 1; the trajectory of body 2 is a distance $s$ away from the XY plane. When two bodies reaches the minimum distance $\Dmin$, their projected relatiisve distance on the XY plane becomes $\sqrt{ {\Dmin}^2 - s^2 }$.}
\label{DgrmLn}
\end{figure}

\subsection{Extension to Non-intersecting Orbits\label{s:extnintsec}}
Even when two orbits do not intersect, collision is still possible if the distance between the closest points of the two orbits, so-called ``Minimum orbit intersection distance'' or MOID, is less than the collision radius $\tau$.  We note that, in general there exist multiple local minima of the distance between two orbits \citep{Gronchi:2005}, and it is possible that more than one of these could be less than $\tau$. 
The collision probabilities introduced by those local minima can be easily integrated, if necessary, as they have the same functional form as that of the MOID.  In the case study that we describe in Section~\ref{s:casestudy}, we include contributions by all the multiple local minima.

The calculation of $P$ for the non-intersecting case and its averaging for MOID$<\tau$ is described well in \citet{Greenberg:1982}.  Here we derive the same equation with an alternative approach. 
First we note that two skewed lines, {\it i.e.} a pair of lines not intersecting nor being parallel, have a unique minimum distance and the vector along the minimum distance is normal to both the lines. This can be generalized as the following theorem.
\begin{theorem}
\label{t:skewd}
Two smooth curves have the local minimum distance along the line orthogonal to the tangents of both curves.
\end{theorem}
Stated in other words, the normal planes of two smooth curves at a certain point on each curve should coincide with each other if there exists a local minimum distance at the given points.

Then, as before, we linearize the motion of the two bodies in the neighborhood of the MOID. Figure~\ref{DgrmLn}  illustrates the situation when the  minimum non-zero distance between two orbits, denoted by ${\bf s}$, is along the Z axis.  By Theorem~\ref{t:skewd}, both bodies move normal to the Z axis in the vicinity of the MOID location. Without loss of generality, we choose the Y axis to be parallel to the motion of body 2. The path of body 2, shifted by $- {\bf s=-(0,0,s)}$, intersects the path of body 1 at the origin, as in Section~\ref{s:drvintsec}. 
Therefore, from Equation~\ref{rtv} and~\ref{rXvrel}, 
\begin{equation}
{\bf r_1} + t_1 {\bf v_1} = {\bf r_2} - {\bf s} + t_2 {\bf v_2},
\label{rtv2}
\end{equation}
\begin{equation}
({\bf r_1} - {\bf r_{2}} +  {\bf s} ) \times {\bf U} = (t_1 - t_2 ) ({\bf v_1} \times {\bf v_2} ).
\label{rXvrel2}
\end{equation}
As before, we consider body 1 and body 2 to have a minimum distance $\Dmin$ at ${\bf r_1}$ and ${\bf r_2}$ at $t=0$.
In Equation~\ref{rXvrel2}, the resultant direction of the right-hand side is parallel to the Z axis. As ${\bf U}$ is on the XY plane, ${\bf r_1} - {\bf r_{2}} +  {\bf s}$ should also lie on the XY plane and its magnitude is 
\begin{equation}
\left| {\bf r_1} - {\bf r_{2}} +  {\bf s} \right|  = \sqrt{ {\Dmin}^2 - s^2 }.
\end{equation}
from Pythagorean theorem (Figure~\ref{DgrmLn}). Interestingly, ${\bf U}$ is normal to both ${\bf s}$ and ${\bf r_1} - {\bf r_2}$; the latter is normal to ${\bf U}$ according to Theorem~\ref{t:normal}. Therefore, ${\bf r_1} - {\bf r_{2}} +  {\bf s}$ is also normal to ${\bf U}$, which gives
\begin{equation}
\Delta t_{\rm col} = \frac{\tau U }{\left| {\bf v_1} \times {\bf v_2} \right|} \sqrt{ 1 - \frac{s^2}{\tau^2}}.
\label{Dt2}
\end{equation}
for collision radius $\tau$.

For two fixed orbits having $0<s \leq \tau$, we can use Equation~\ref{Dt2} without any modification. However, in Monte-Carlo type numerical simulations, the averaged value of $P$, within $0<s \leq \tau$, is more useful as we can get a better estimate of impact flux from the same number of test projectiles.

In the small vicinity where collisions are allowed, $s$ can be assumed to be uniformly distributed.  (This follows from the argument that, for a random distribution of lines with fixed directions in space, the fraction of cases with minimum distance less than $s$ is proportional to $s$; this is a good assumption for $\tau\ll$ heliocentric distance.) Averaging $\Delta t_{\rm col}$ within  $0<s \leq \tau$, Equation~\ref{Dt2} gives $\pi / 4$ times $\Delta t_{\rm col}$ of Equation~\ref{Dtcol}. Therefore, the averaged collision probability per unit time when two bodies have $s$ random in the range $(0,\tau)$ becomes
\begin{equation}
P = \frac{\pi \tau U }{2 \left| {\bf v_1} \times {\bf v_2} \right| T_1 T_2}.
\label{Ptau}
\end{equation}

\subsection{Equivalence with Wetherill's Expression\label{s:eqv}}
In \citet{Wetherill:1967}'s derivation, his Equation 7 gives the probability of collision per unit time for two bodies on intersecting orbits.  Expressing his equation with our notation gives
\begin{equation}
P  = \frac{\pi \eta_1}{2 \left| {\bf v_1} \right| T_1 T_2},
\label{WethPX}
\end{equation}
where 
\begin{equation}
\eta_1 = \frac{\tau U}{\sqrt{U^2 - { \left(  U_x  \cos{\alpha_1}+ U_y \sin{\alpha_1} \right)^2}}}.
\label{etamax}
\end{equation}
In above equation $U_x$ and $U_y$ are the X and Y components of ${\bf U}$ and $\alpha_1$ is the angle between the X axis and the velocity vector of body 1. The Sun is located in the --X direction. The physical meaning of $\eta_1$ is the maximum distance from the orbit intersection point where body 1 should be present to allow collision to occur when body 2 is at the orbit intersection point. Derivation of Equation~\ref{WethPX} and \ref{etamax} is lengthy and complicated in \citet{Wetherill:1967} and his equation does not look commutative between body 1 and body 2 at first glance. Below we prove that his equation is the same as our simplified derivation of the collision probability per unit time, Equation~\ref{Ptau}, which is clearly symmetric in body 1 and body 2.

The term,  $U_x  \cos{\alpha_1}+ U_y \sin{\alpha_1}$, in the denominator of Equation~\ref{etamax} is the projection of the {\bf encounter} velocity ${\bf U}$ along the velocity direction of body 1. Therefore, the denominator is equal to the component of  ${\bf U}$ normal to ${\bf v_1}$. Thus,
\begin{equation}
\eta_1 = \frac{\tau U \left| {\bf v_1} \right| }{\left|  {\bf U} \times {\bf v_1} \right|} = \frac{\tau U \left| {\bf v_1} \right| }{\left| {\bf v_1} \times {\bf v_2} \right|},
\end{equation}
which gives Equation~\ref{Ptau} when it is substituted to Equation~\ref{WethPX}.

\section{Collision Probability for Tangential Encounters}\label{s:tangenc}

\subsection{Derivation for Intersecting Orbits\label{s:drvintsec2}}

\begin{figure}
\centering
  \includegraphics[width=300px]{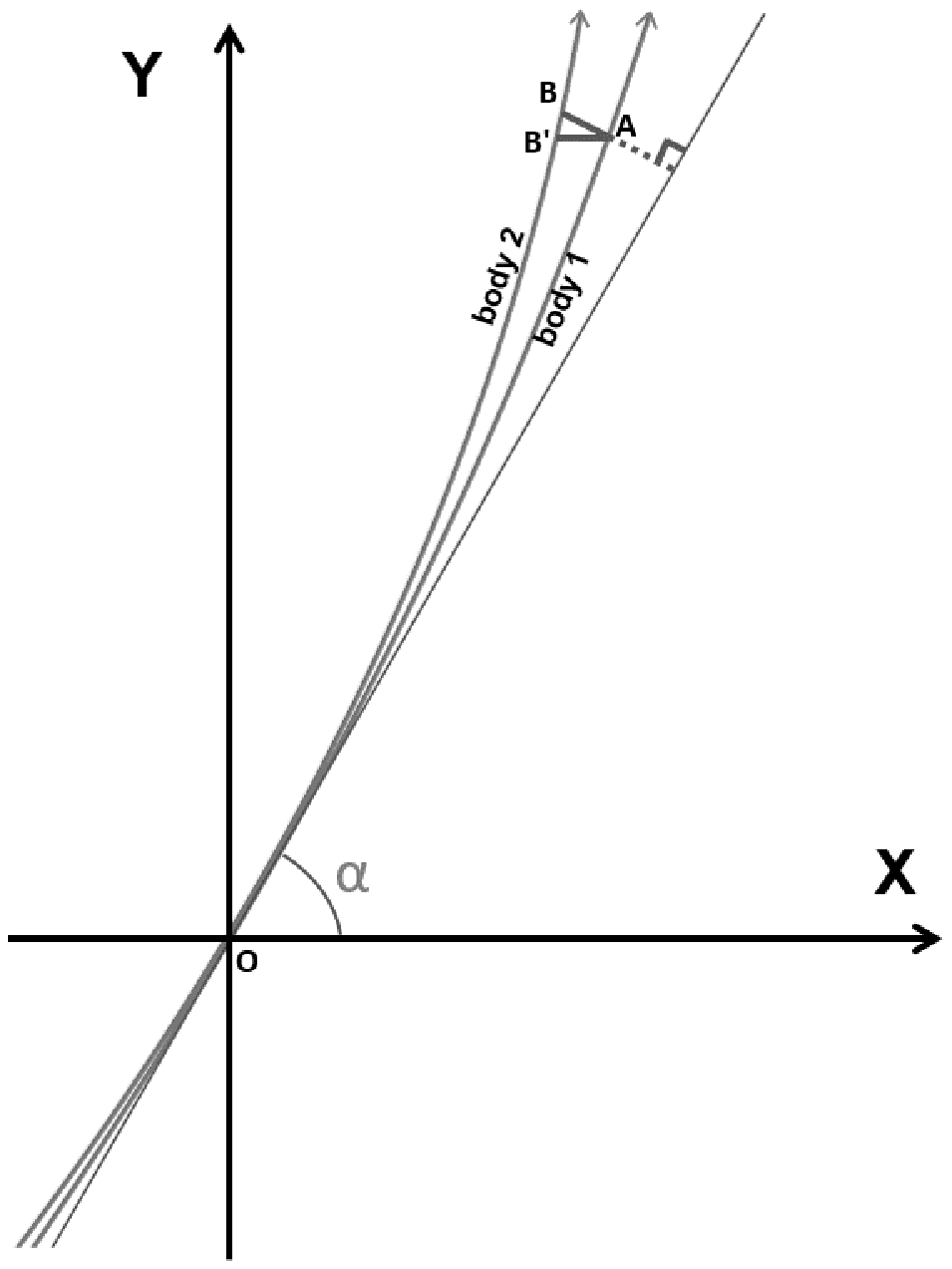}
  \caption{Diagram of two tangentially intersecting orbits. The Sun is located along the minus horizontal (-X) direction and the orbit intersection is located at the origin. The minimum distance between body 1 and body 2 is $\Dmin$ when the bodies are located at point A and B, respectively. Point B$'$ is the location of body 2 having the same Y component as point A. The angle between the common direction of two bodies (staight line) and +X direction is $\alpha$. The curvatures of two bodies are exaggerated for visual clarification.}
\label{DgrmTng}
\end{figure}

It is evident that the right hand side of Equation~\ref{Ptau} is singular when ${\bf v_1}$ and ${\bf v_2}$ are parallel. This situation is illustrated in Figure~\ref{DgrmTng}.  The origin of this singularity lies in our linear approximation for the motion of bodies near the collision point. To resolve this singularity, we can use a better approximation of the true motion of the two bodies, namely parabolic motion instead of linear motion,
\begin{equation}
{\boldsymbol \rho}_i (t) =  {\bf r}_i + t {\bf v}_i + \half (t-t_i )^2 {\bf g} - \half {t_i}^2 {\bf g},
\end{equation}
where ${\bf g}$ is the gravitational acceleration due to the Sun, assumed to be constant in the vicinity of the encounter. The constant vector ${\bf v}_i$ is the velocity of body $i$ at the orbit intersection point, and body $i$ passes the orbit intersection point at time $t_i$. At this time, the direction of the velocity vector of both bodies is the same.  The encounter geometry is illustrated in Figure~\ref{DgrmTng} in a coordinate system with the origin at the orbit intersection point, the Sun is located on the negative X axis, the orbital poles of the two bodies are aligned to the Z direction, and the motion of the two bodies is approximated as parabolic paths.  (Note that the time epochs, $t_1$ and $t_2$, when body 1 and body 2 pass the origin, have negative values in the case illustrated in Figure~\ref{DgrmTng}.) We denote by $\alpha$ the angle between the X-axis and the common direction of the velocity vectors; the range of $\alpha$ is $0$ to $\pi$. The time $t=0$ is defined as the epoch when the two bodies approach the minimum distance, $\Dmin$; at this time their position vectors are ${\bf r}_1$ and $ {\bf r}_2$ and the encounter velocity is given by the relative velocity at $t=0$, 
\begin{equation}
{\bf U_0} = {\bf v_1} - {\bf v_2} - (t_1 - t_2) {\bf g}.
\label{U0}
\end{equation}

We define the outer orbit, the one having smaller curvature, as body 1 and the inner orbit as body 2. Because the outer body should have higher velocity than the inner body at  $t=0$, we introduce a velocity ratio 
\begin{equation}
k = {v_2} / {v_1}, \qquad -1 < k < 1,
\label{e:k}
\end{equation} 
where the negative values occur when body 2 orbits in the opposite direction of body 1. For simplicity, in the following we consider only the case of positive $k$ for the derivation of $P$. The derivation for the negative $k$ is similar, and we note that the final results (Equation~\ref{tcoltng}, \ref{Ptng0}, \ref{DtTng2}, and \ref{Ptng}) hold true for both positive and negative $k$.  The general derivation with an alternative approach is also provided in the Appendix.

The path of body 2 from the origin to point B is slightly longer than the path of body 1 from the origin to point A in the case illustrated in Figure~\ref{DgrmTng}.  Because the Y components of the velocities remain constant, the ratio of the travel time for $\overline{OB'}$ for body 2 and $\overline{OA}$ for body 1 is equal to the inverse of the ratio of initial velocity, $1/k > 1$. (Here $B' $ is the location of body 2 having the same Y component as the point A.)   The relation between the times $t_1$ and $t_2$ (when body 1 and body 2 pass the origin)  can be written as
\begin{equation}
t_2 = \frac{t_1}{k} - \delta t,
\label{t2t1delt}
\end{equation}
where the small time interval, $\delta t$, for body 2 to go from $B'$ to $B$ is given by
\begin{equation}
\delta t \simeq \frac{ \Dmin }{ v_2 \tan{\alpha} }.
\label{delt}
\end{equation}   
Note that $\delta t$ is positive when $0< \alpha < \pi/2$ and negative for $\pi/2< \alpha < \pi$.  

To solve for $t_1$ and $t_2$, we take a similar approach as in Section~\ref{s:nontngcol}.  Thus, similar to Equation~\ref{rtv}, we can set up the following relation for the positions of the two bodies using the orbit intersection point:
\begin{equation}
{\bf r_1} + t_1 {\bf v_1} - \frac{1}{2} {t_1}^2 {\bf g} = {\bf r_2} + t_2 {\bf v_2} - \frac{1}{2} {t_2}^2 {\bf g}.
\label{rtv3}
\end{equation}   
By taking the vector product with the encounter velocity, ${\bf U}_0$ (Eq.~\ref{U0}), on both sides of Equation~\ref{rtv3}, and using ${\bf v_1} \times {\bf v_2}=0$ (for tangential encounters), we get
\begin{equation}
({\bf r_1} - {\bf r_2} ) \times {\bf U_0} = \frac{1}{2} ( {t_1}^2 - {t_2}^2 ) {\bf g} \times {\bf U} +  (t_1 - t_2 ) {\bf g} \times (t_2 {\bf v_2} - t_1 {\bf v_1}).
\label{rXvrel3}
\end{equation}   
where ${\bf U}={\bf v_1} - {\bf v_2}$.  With the use of Theorem~\ref{t:normal}, the above equation simplifies to the following,
\begin{equation}
\Dmin U_0 = \left[ \frac{1}{2} ( {t_2}^2 - {t_1}^2 ) U +  (t_2 - t_1 )  \delta t \, v_1  \right] g \sin{\alpha},
\label{rXvrel3_2}
\end{equation}
Because $\delta t$ is a small value, we neglect the term with $\delta t$ for the moment (but we return to it below). 
Also, considering that $\left| t_1 - t_2 \right| g$ is much smaller than $\left| {\bf v_1} - {\bf v_2} \right|$, ${\bf U}$ can be approximated as ${\bf U_0}$. Then, we find the minimum distance of approach of the two bodies is related to their times of passage at the origin as follows,
\begin{equation}
\Dmin = \frac{1}{2} ( {t_2}^2 - {t_1}^2 ) g \sin{\alpha}.
\label{dmin}
\end{equation}
By using $t_1\simeq kt_2$ and rearranging Equation~\ref{dmin}, we obtain
\begin{equation}
 t_2 = \mp \sqrt{\frac{2 \Dmin}{(1-k^2 ) g \sin{\alpha}}},
\label{t2}
\end{equation}
where the minus sign is for the case when the bodies have already passed the origin when the minimum distance is achieved (as in Figure~\ref{DgrmTng}), whereas the plus sign is for the opposite case.

Analogous to the derivation in Section~\ref{s:nontngcol}, we can define the time interval $\Delta t_{\rm col} = \left| t_2 - t_1 \right|$ for collision to occur for a given collision radius $\tau$:
\begin{equation}
\Delta t_{\rm col} = \sqrt{\frac{2 (1-k) \tau }{(1+k) g \sin{\alpha} }}.
\label{tcoltng}
\end{equation}
Thus, if body 2 has passed the origin more than $\Delta t_{\rm col}$ earlier than body 1, body 1 cannot catch up to collide with body 2. On the other hand, unlike Figure~\ref{DgrmTng}, in order for the collision to occur before the bodies reach the orbit intersection point, the expected $t_1$ should not be more than $\Delta t_{\rm col}$ earlier than the expected $t_2$.

If we wish to be more accurate, Equations~\ref{t2t1delt} and \ref{dmin} give the following better estimate of $\Delta t_{\rm col}$,
\begin{equation}
\Delta t_{\rm col} \simeq \sqrt{\frac{2 (1-k) \tau }{(1+k) g \sin{\alpha}}} \pm \frac{k \delta t}{1+k},
\label{tcoltng2}
\end{equation}  
where the sign order is the same as in Equation~\ref{t2}.

In other words, regardless of the choice between the above two equations, if body 2 passes the origin within a time interval $\sim 2 \Delta t_{\rm col}$ of when body 1 passes the origin, the collision would occur. (Note that the $\delta t$ term in Equation~\ref{tcoltng2} cancels when we add the time intervals with the plus and minus signs.) Thus, the collision probability per unit time of these two bodies whose velocity vectors are parallel near the collision point is given by
\begin{figure}
\centering
  \includegraphics[width=300px]{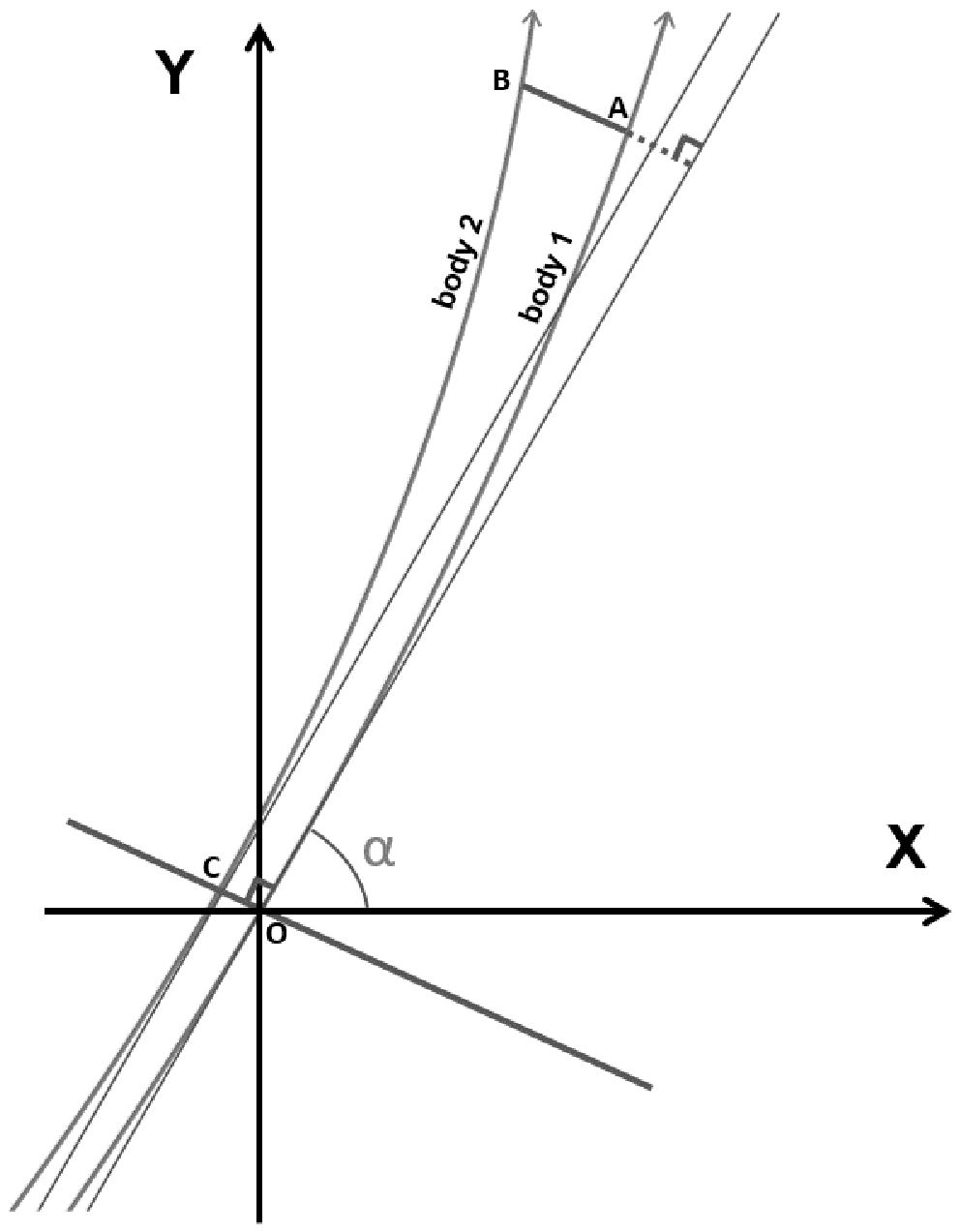}
  \caption{Similar diagram as Figure~\ref{DgrmTng}. The Sun is located along the minus horizontal direction (-X) and the path of body 1 (OA) defines the XY plane. The projection of body 2 on the XY plane is also shown (OB). Minimum orbit intersection occurs when body 1 is at the origin (OC). The minimum distance with time between two orbit $\Dmin$ is projected on XY plane and shown as line AB. The moving direction of two bodies at A and B are almost same as their direction at the origin (two straight lines). However, we exaggerated curvatures of two bodies for visual clarification.}
\label{DgrmTng2}
\end{figure}
\begin{equation}
P = \frac{2 \Delta t_{\rm col}}{T_1 T_2} = \frac{1}{T_1 T_2} \sqrt{\frac{8 (1-k) \tau }{(1+k) g \sin{\alpha}}}.
\label{Ptng0}
\end{equation}
This equation may appear counter-intuitive at first glance because $P$ decreases as the velocity ratio $k$ approaches 1. It is true that the path length OA and OB increases as $k$ approaches 1. However, as the two velocities become identical, it takes longer for body 1 to catch up with body 2. Consequently, when body 1 is at the origin, body 2 should be located within a smaller distance to have a chance to collide with body 1. 

\subsection{Extension to Non-intersecting Orbits\label{s:extnintsec2}}
Now we consider a more general collision geometry where the orbits do not intersect but their minimum approach distance, $s$, is small enough to allow collision. We define a coordinate system that has the origin as the point on  body 1's orbit where the MOID, $0<s \leq \tau$, occurs. Again, the Sun is located along the --X direction and the orbit of body 1 is on the XY plane. We assume that the variation of the vertical velocity component of body 2 is negligible in the small region where collisions are possible. 

According to Theorem~\ref{t:skewd}, at the location where MOID occurs, the position vector of body 2 should be normal to the velocity direction of both bodies, $(\cos{\alpha},\sin{\alpha},0)$. Thus, the position of body 2 at $t=t_2$ can be expressed as follows:
\begin{equation}
{\bf s} = s ( -\cos{\beta} \sin{\alpha}, \cos{\beta} \cos{\alpha}, \sin{\beta}).
\end{equation}
where $\beta$ is the angle between the position vector of body 2 and the XY plane.

\begin{figure}
\centering
  \includegraphics[width=300px]{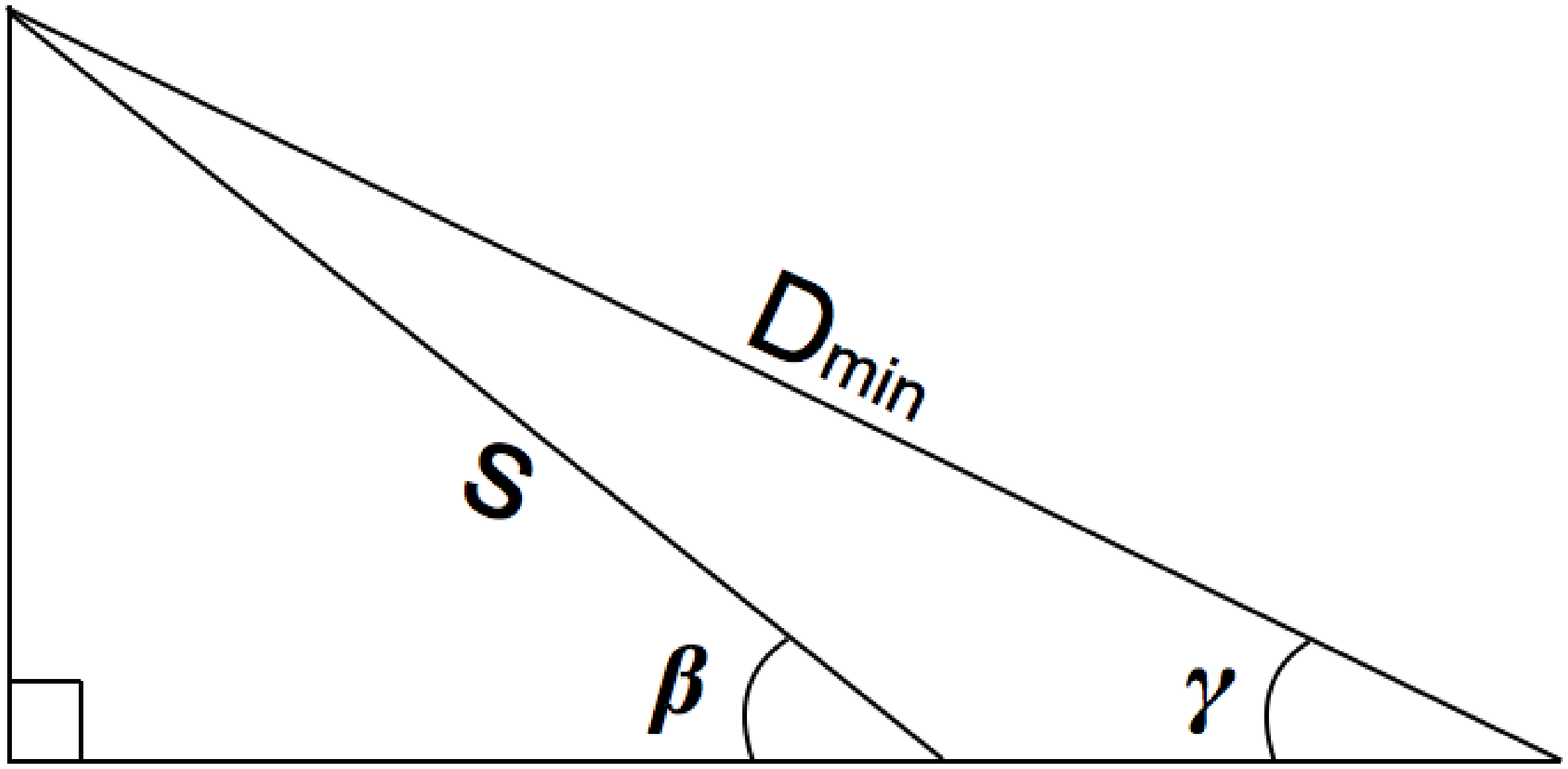}
  \caption{Schematic diagram depicting the geometry of the MOID $s$, minimum distance of two moving bodies $\Dmin$, and the angles $\beta$ and $\gamma$ measured from the XY plane.}
\label{DgmSbta}
\end{figure}

Figure~\ref{DgrmTng2} shows the orbit of body 1 on the XY plane with the projection of the orbit of body 2 on the XY plane. As before, when body 1 passes the point A, it can barely touch body 2 located at B. Because the path of body 2 shifted by $- {\bf s}$ intersects the path of body 1 at the origin, from Equations~\ref{rtv3} and \ref{rXvrel3}, we get
\begin{equation}
{\bf r_1} + t_1 {\bf v_1} - \frac{1}{2} {t_1}^2 {\bf g} = {\bf r_2} - {\bf s} + t_2 {\bf v_2} - \frac{1}{2} {t_2}^2 {\bf g},
\label{rtv4}
\end{equation}
\begin{equation}
({\bf r_1} - {\bf r_2} + {\bf s}) \times {\bf U_0} = \frac{1}{2} ( {t_1}^2 - {t_2}^2 ) {\bf g} \times {\bf U} +  (t_1 - t_2 ) {\bf g} \times (t_2 {\bf v_2} - t_1 {\bf v_1}).
\label{rXvrel4}
\end{equation}
As in Section~\ref{s:drvintsec2}, we approximate ${\bf U_0} \simeq {\bf U}$ and $t_2 \simeq k t_1$. As ${\bf r_1} - {\bf r_2} \perp {\bf U_0}$ (Theorem~\ref{t:normal}) and ${\bf s} \perp {\bf U}$ (Theorem~\ref{t:skewd}), ${\bf r_1} - {\bf r_2} + {\bf s}$ is also close to orthogonal to ${\bf U_0}$ and ${\bf U}$, therefore we get
\begin{equation}
\left| {\bf r_1} - {\bf r_2} + {\bf s} \right|  = \frac{1}{2} ( {t_2}^2 - {t_1}^2 ) g \sin{\alpha}.
\label{dmin2}
\end{equation} 
The Z-directional motion of body 2 is negligible in the small vicinity of the MOID location, so body 2 is moving in the plane parallel to the orbital plane of body 1.  Thus, as illustrated in Figure~\ref{DgmSbta}, ${\bf r_1} - {\bf r_2} + {\bf s}$ lies in the XY plane and has length
\begin{equation}
\left| {\bf r_2} - {\bf r_1} - {\bf s} \right| = \Dmin \cos{\gamma} - s \cos{\beta},
\end{equation} 
where $\gamma$ is the angle between the line corresponding to the MOID and the XY plane. By the law of sines, the angle $\gamma$ is given by
\begin{equation}
\Dmin \sin{\gamma} = s \sin{\beta}.
\label{Dgmeqsbta}
\end{equation}
As is the case in Section~\ref{s:drvintsec2}, $t_2$ is not exactly the same as $t_1 / k$, but the difference almost cancels with the  time difference on the opposite side, {\it i.e.} the case where the minimum distance is achieved before the bodies reach the origin. Thus, from Equation~\ref{dmin2} through~\ref{Dgmeqsbta}, we obtain, for collision radius of $\tau$, the time interval $\Delta t_{\rm col}$ to be
\begin{equation}
\Delta t_{\rm col} = \sqrt{ \frac{2 (1-k) \tau}{(1+k) g \sin{\alpha}}}  \left( \sqrt{1 -  \frac{s^2}{\tau^2} \sin^2{\beta}}  - \frac{s}{\tau} \cos{\beta}  \right)^{1/2} .
\label{DtTng2}
\end{equation}

Assuming a uniform distribution within the parameter space $0<s \le \tau$ and $- \pi /2 \le \beta \le \pi/2$, we can obtain the average value of $\Delta t_{\rm col}$.
(Care must be taken over the range of $\beta$, as the orbit of the faster body should always be outside of the slower body in the vicinity of the MOID.) With numerical integration we found the average value of the factor $( \sqrt{1 -  \frac{s^2}{\tau^2} \sin^2{\beta}}  - \frac{s}{\tau} \cos{\beta})^{1/2}$ is 0.61. Thus, we obtain the collision probability per unit time,

\begin{equation}
P = \frac{1.7}{T_1 T_2} \sqrt{ \frac{(1-k) \tau}{(1+k) g \sin{\alpha}}} .
\label{Ptng}
\end{equation}

In the Appendix, we present fully analytic derivations which prove that Equation~\ref{tcoltng2} and \ref{DtTng2} are valid for $\tau / r \ll 1-k$. This condition is well satisfied in solar system applications of the impact rate of asteroids on planets, since planet sizes are much smaller than their heliocentric distances.  An example where this condition may be violated is for large planets orbiting close to their host star.  For example, in the case of hot jupiter WASP-18b orbiting with semimajor axis of 0.02~AU, the physical radius is 2.7\% of the semimajor axis \citep{Southworth:2009}. Considering the gravitational focusing factor, we calculate $\tau / r$ of WASP-18b to be equal to $1-k$ when $k=0.7$. For the more common planetary targets with larger semimajor axis and smaller radius than WASP-18b, Equation~\ref{tcoltng2} and \ref{DtTng2} are valid approximations unless the encounter velocity is vanishingly small.

\section{Transition between Non-tangential and Tangential Encounters\label{s:transition}}
We have seen that the collision probability for non-tangential encounters, Equation~\ref{Pmoid0} and \ref{Ptau}, increases to infinity as the velocity vectors become aligned, which is not physical. On the other hand, our formulas for the collision probability for the tangential case, Equation~\ref{Ptng0} and \ref{Ptng}, are independent of the angle, $\theta=\arccos({\bf v_1\cdot v_2}/v_1v_2)$, between the two velocity vectors. This is also not a good approximation when $\theta$ is not too small. 
We expect that the collision probability smoothly transitions from the non-tangential formula to the tangential case as $\theta$ decreases 
below some transition angle, $\theta_c$. 
However, a rigorous calculation of this transition is hard to obtain, for two reasons. First, the collision time interval $\Delta t_{\rm col}$ (Equation~\ref{Dt2} and \ref{DtTng2}) is a  function of $s$ and $\beta$, therefore, the transition will be a function of both parameters.  Furthermore, for tangential collisions, the body with higher velocity should orbit outside of the  body of slower velocity near the MOID location, whereas this restriction is removed for the non-tangential case.

Here, we provide an approximate condition for the transition between the non-tangential and tangential collisions by equating Equation~\ref{Ptau} and \ref{Ptng}. When the velocity vectors are nearly parallel, we have $\V_2\simeq k\V_1$ and $U\simeq(1-k)v_1$.  Then, with the approximation $\sin{\theta} \simeq \theta$, we obtain the transition value $\theta_c$,
\begin{equation}
\theta_c \simeq 0.9 {\sqrt{(1-k^2 ) \tau g \sin{\alpha}} \over {k v_1}}.
\label{thetacrit2}
\end{equation}
It is useful to comment on some properties of $\theta_c$. 

\begin{enumerate}

\item The $\sin{\alpha}$ term in the numerator can be expressed as a function of true anomaly $f$ and eccentricity $e$ of an elliptical orbit,
\begin{equation}
\sin{\alpha} = {1+e\cos{f} \over \sqrt{1+ 2e\cos{f} +e^2 }}.
\label{sinalpha}
\end{equation}
The minimum value of $\sin{\alpha}$ is $\sqrt{1-e^2}$, and it occurs when $\cos{f}=-e$.  For nearly circular orbits, $\sin{\alpha}$ is close to unity.  For an eccentric orbit of $e=0.5$, the minimum $\sin{\alpha}$ is 0.866 when $f=\pm 120^\circ$, and $\sin\alpha$ can be as small as 0.5 only when the eccentricity is as large as 0.866.

\item The factor $\sqrt{1-k^2} / k$ is monotonically decreasing to zero as $k$ approaches unity.  This means that $\theta_c$ is smaller for smaller encounter velocities.

\item The transition angle, $\theta_c$, is proportional to the square root of the collision radius divided by the heliocentric distance, $\sqrt{\tau/r}$. Therefore, the transition angle would be significantly larger for close-in large planets, such as hot Jupiters in exo-planetary systems. 
 
 \end{enumerate}

\begin{figure}
\centering
  \includegraphics[width=300px]{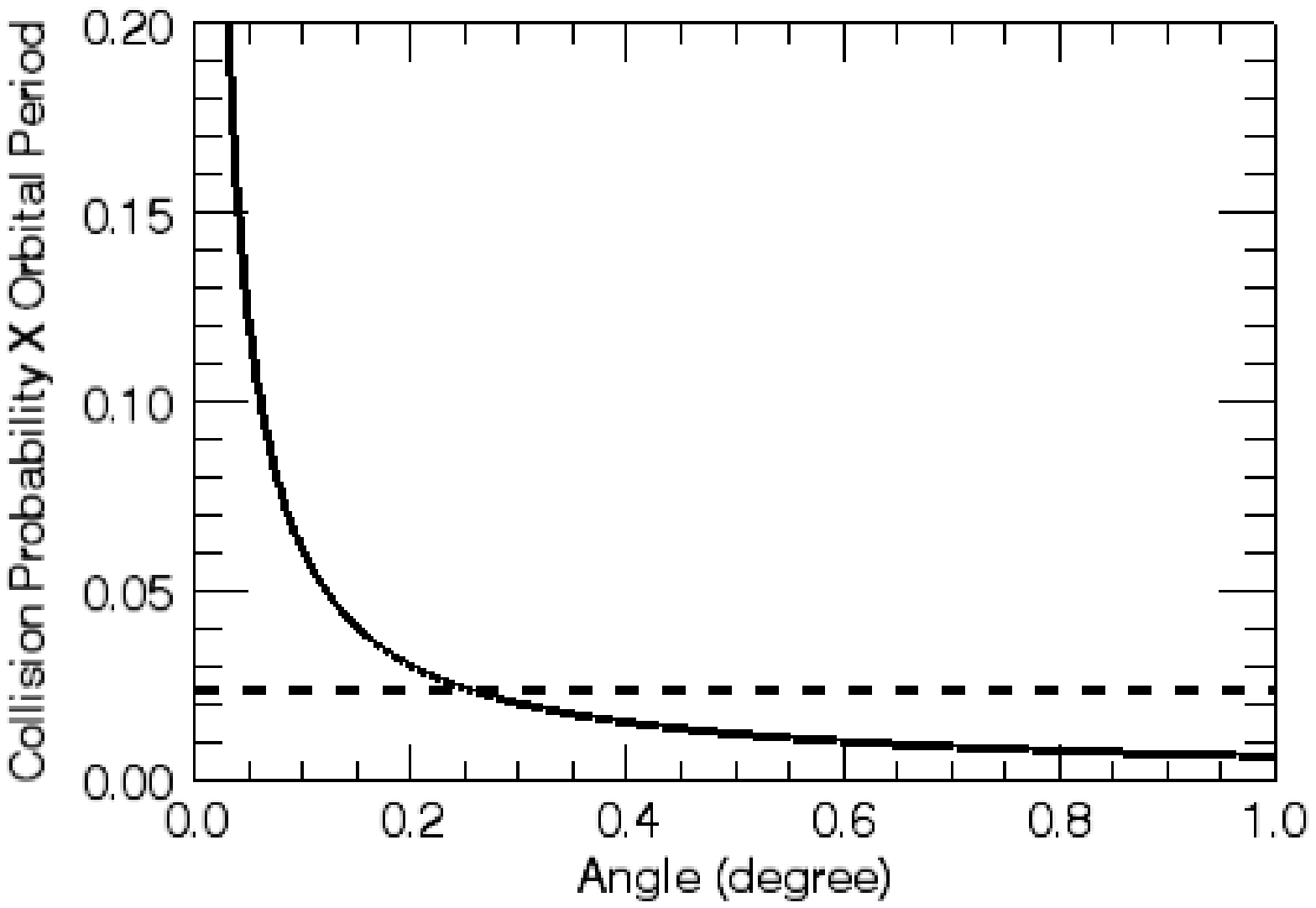}
  \caption{Collision probability with Earth of a small body having orbital velocity of $0.8v_\oplus$ near the MOID location. Earth's orbit is assumed to be circular and gravitational focusing is neglected. The ordinate is the collision probability per orbital revolution of Earth. The previous method of calculation has an unphysical singularity near $\theta=0$ (solid line, Equation~\ref{Ptau}); our new method for calculation of collision probability of a tangential encounter gives a finite result which is independent of the encounter angle (dashed line, Equation~\ref{Ptng}). }
\label{ColPcomp}
\end{figure}

As an example, consider the case of the Earth with a circular orbit and an impactor with velocity ratio of $k=0.8$.  We adopt the physical radius of the Earth for $\tau$ (neglecting gravitational focusing). Figure~\ref{ColPcomp} shows the collision probability $P$ for this example, calculated with Equations~\ref{Ptau} and \ref{Ptng} (solid and dashed lines, respectively). (We plot the product of $P$ and the orbital period of the impactor to provide the dimensionless result, collision probability per orbital revolution of Earth.)  For this case, we find $\theta_c \simeq 0.26^\circ$. 

In practical Monte-Carlo type numerical simulations with a moderate number of particles, the number of cases of MOID $< \tau$ is often statistically too small to accurately calculate the integrated impact flux of a projectile population on a given target. For computational efficiency, an artificially enhanced collision radius, $\tau'=p\tau$ with $p\gg1$, can be adopted and the results scaled to the real collision radius \citep[e.g.][]{JeongAhn:2015}. For the case of non-tangential encounters, the number of random orbits having MOID $< \tau$ on the target orbit increases linearly with $\tau$. As the collision probability itself also linearly increases with $\tau$ (Equation~\ref{Ptau}), the total impact flux is proportional to $\tau^2$.  Thus, the numerically computed impact flux with an adopted collision radius $\tau'=p\tau$ can be rescaled by a factor $p^{-2}$ to obtain the real impact flux. 

On the other hand, for the case of tangential encounters, the number of cases of MOID $<\tau$ is proportional to $\tau^2$.  In practice, however, the exactly tangential cases are of measure zero, so two velocity vectors will not be strictly parallel near the MOID location, and there would be a small, non-zero angle between them.  We can identify the near-tangential cases, i.e., those with $\theta<\theta_c$, and in these cases the minimum encounter distance should be measured along the $\V_1 \times \V_2$ direction. The number of such cases having MOID $< \tau$ is proportional to $\tau$.  As the collision probability in these encounters is proportional to $\sqrt{\tau}$ (Equation~\ref{DtTng2}), the integrated impact flux of these cases is proportional to $\tau^{3/2}$.   Therefore, in Monte-Carlo type numerical simulations, care should be taken with the artificially enhanced collision radius: the transition angle $\theta_c$ should be calculated with the physical radius $\tau$ (including a gravitational focusing factor), not the artificially enhanced collision radius; and the impact flux of the near-tangential collisions should be rescaled by a factor $p^{-3/2}$.

\section{Case Study}\label{s:casestudy}
In this section, we provide a case study of the impact flux of a synthetic population of NEOs on Earth to illustrate that the correct treatment of tangential encounters is crucial to determine the correct impact frequency. The orbital parameters, the mass and the physical radius of Earth are adopted and the impact frequency from $N=5\times10^6$ test particles orbiting in Earth-like orbits is calculated.  The test particles' semimajor axes are chosen randomly from the range 1.1--1.2~\AU; this range exceeds ten times the Hill radius of Earth, and avoids the co-orbital region of Earth. The eccentricities are chosen randomly from zero to 0.3, and inclinations from zero to 5 degrees relative to the ecliptic; these ranges are selected for more frequent near-tangential encounters. The arguments of perihelia and the longitudes of ascending node are randomly generated within the range zero to $2\pi$.  For statistical analysis, we repeat the same simulation 100 times (making different realizations of the random orbital parameters in each case), and we report mean values and standard deviations of the impact counts below.

Using the code developed by \citet{Gronchi:2005}, we determine that the total number of orbit crossings having local minimum distances smaller than the collision radius is $39019\pm220$. For the collision radius, we multiply the physical radius of Earth by the gravitational focusing factor of each test particle.  For this synthetic population of impactors with Earth-like orbits, the mean of the gravitational focusing factor is a significantly large value of 2.96.

\begin{figure}
\centering
  \includegraphics[width=400px]{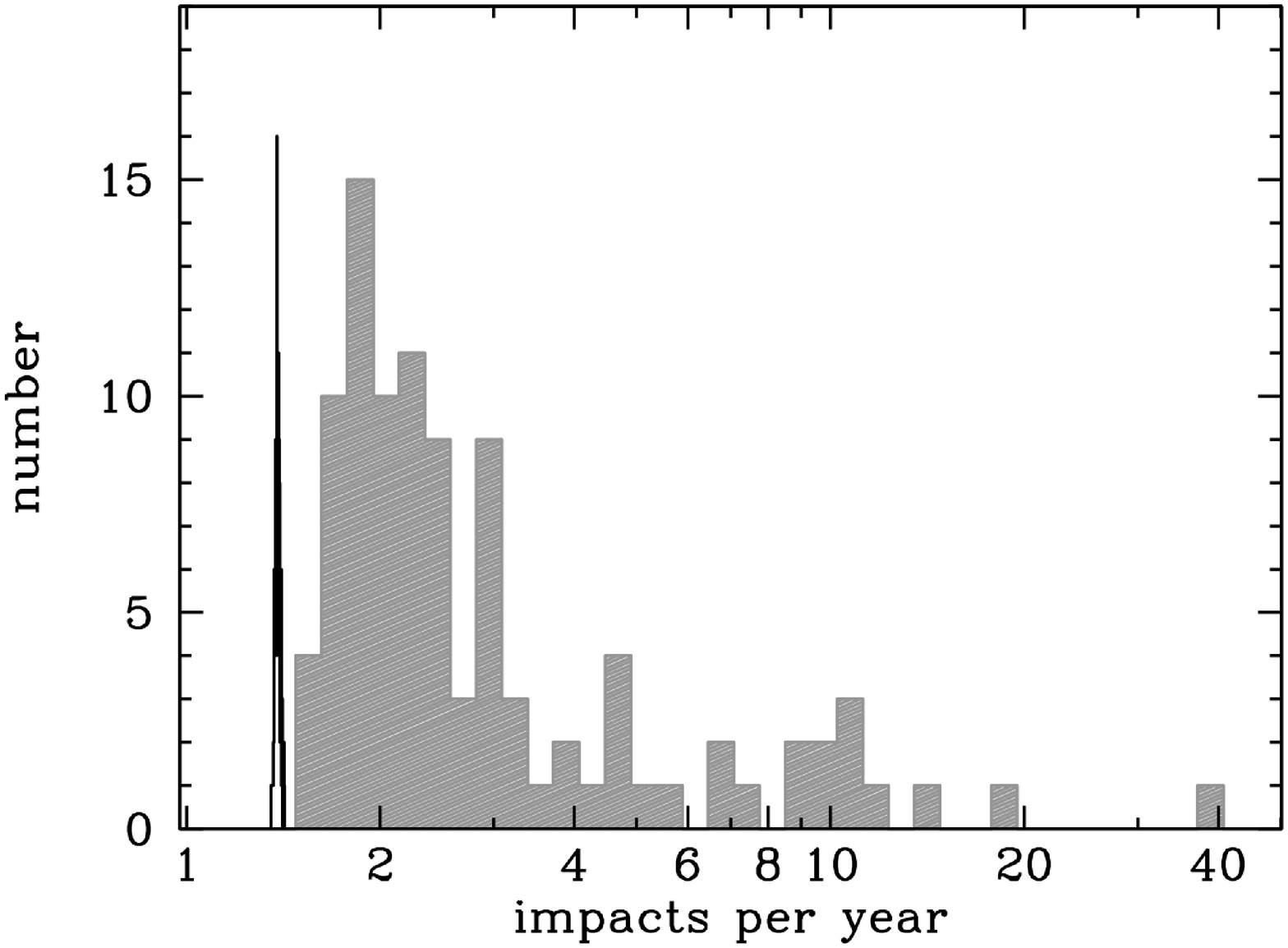}
  \caption{Distribution of the impact rates on Earth from a synthetic population of $5\times10^6$ test particles representing NEOs.  The gray shaded histogram (with wide bins) depicts the results from the unmodified \"Opik-Wetherill formulas using Equation~\ref{Ptau}.  The black histogram (with narrow bins) depicts the results from our improved method in which Equation~\ref{Ptau} is replaced by Equation~\ref{Ptng} for the near-tangential encounters. The impact rates from 100 independent realizations of random test particle NEOs are used for these histograms. Note that a logarithmic scale is used on the abscissa.}
\label{ImpFrqLog}
\end{figure}

The total impact frequency calculated from Equation~\ref{Ptau} varies greatly from simulation to simulation, from a minimum of 1.6 impacts per year to the maximum of 456 impacts per year;  the results are shown in the gray shaded histogram in Figure~\ref{ImpFrqLog}.  The mean value and the standard deviation are 9.8 and 47, respectively. This highly variable impact frequency is owed to the divergence of Equation~\ref{Ptau} when test particles sporadically encounter Earth at nearly tangential velocities with $\theta \simeq 0$. A careful examination of the distribution of the close encounters reveals that, among the $39019\pm220$ local minimum distances smaller than the collision radius, only $50\pm8$ of them have smaller $\theta$ than the critical value (Equation~\ref{thetacrit2}), but the smallest $\theta$ case alone contributes more than 50 percent of the total impact frequency in 18 simulations out of 100. However, the corrected total impact frequency, when Equation~\ref{Ptau} is replaced by Equation~\ref{Ptng} for the near-tangential encounters, is found to be a very stable quantity, $1.39\pm0.01$ impacts per year (shown in the narrow black histogram on the left of Figure~\ref{ImpFrqLog}).  These results show that the incorrect treatment of near-tangential encounters leads to systematically higher impact rate estimates and also greater scatter of the estimates.

In order to demonstrate the validity of our formulas for the impact probability in near-tangential encounters (Equation~\ref{Ptng}), we sampled 1000 test particle orbits having $\theta < \theta_c$, from amongst the above-described synthetic populations of NEOs.  We numerically integrated these particles with SWIFT-RMVS3 (\url{http://www.boulder.swri.edu/~hal/swift.html}), in a model including the Sun and the Earth with their actual mass and radius. The mean anomalies of the test particles were randomly generated in 10 different realizations, and 10 different numerical simulations are carried out for statistical analysis. We limited the orbit integration time to 10 years to avoid orbit evolution of the particles, and we adopted an integration step size of 1.4 minutes to accurately resolve collisions. From the 10 different numerical integrations, we find an average of $10.4\pm2.4$ impacts.  
We can compare this result of the numerical integrations with the expected number of impacts calculated with the unmodified \"Opik-Wetherill formulas and with our corrected formulas.  For the former, the expected value calculated from Equation~\ref{Ptau} is 1802 impacts over ten years; this large value, which is unsurprising due to the nature of the singularity in Equation~\ref{Ptau} for near-tangential encounters, is strongly in disagreement with the numerical integration result.  For the latter, the expected value calculated from Equation~\ref{Ptng} is 8 impacts over ten years, which is within 1--$\sigma$ of the numerical integration result. 

\section{Summary}\label{s:summary}

The classical method of \citet{Opik:1951} and \citet{Wetherill:1967} for calculating collision probabilities of pairs of objects in Keplerian orbits has been widely used in many problems in planetary dynamics. In this paper, we have given a simplified derivation of the backbone of these calculations.  Our derivation is easier to understand and to relate to the underlying geometry of collisions of Keplerian orbits. Additionally, our formula for the collision probability per unit time, $P$ (Equation~\ref{Ptau}), is explicitly commutative between the two colliding orbits (in contrast with the \"Opik-Wetherill formulas).

We also derived the collision probability for tangential encounters (Equation~\ref{Ptng}); this regularizes a singularity in the \"Opik-Wetherill formulas. We achieve this regularization by  replacing the linear approximation in the vicinity of the collision point with a parabolic approximation of the true motions of the bodies, but otherwise the derivation is similar to the derivation for the non-tangential encounters.  In the Appendix, we provide an alternate, fully analytic derivation, which additionally identifies the domain of applicability of the regularized collision probability of tangential collisions.  Stated qualitatively, our formulas are valid in the regime in which the collision radius is much smaller than the heliocentric distance and the encounter velocity is not a vanishingly small fraction of the orbital velocity. The quantitative condition is described in detail in the Appendix.

The domains of the non-tangential and near-tangential collisions should be chosen based on the critical angle we derived in Section~\ref{s:transition}. The additional step needed in computing collision rates is not computationally expensive by virtue of the large increase in available computing power since the 1960s. The neglect of near-tangential encounter cases has the potential to lead to erroneous results; we demonstrated this by an exemplary, although extreme, case of the collision rates on Earth of a population of particles in Earth-like orbits.

\bigskip
\acknowledgements
We thank the referee, Davide Farnocchia, for helpful comments that improved this paper.  This work is supported by UNAM-DGAPA-PAPIIT (grant  IN107316).	R.~M. acknowledges funding from NASA (grant NNX14AG93G) and NSF (grant AST-1312498).

\appendix

\section{Appendix: Alternative Derivation for Tangential Collisions}
In this appendix, we present an alternative derivation for $\Delta t_{\rm col}$ for tangential collisions. The notation used is the same as in {Section~\ref{s:tangenc}, with a few exceptions as noted.}

\subsection{Derivations for Intersecting Orbits}
Let us set up a coordinate system with origin at the point of intersection of the two orbits, 
the XY plane is the common orbital plane of the two orbits, 
and the X direction is radially outward from the Sun as in Section~\ref{s:drvintsec}.  
In general, the minimum distance between the center-of-figure of the two bodies is not zero 
and it does not occur at the point of intersection of the two orbits.
Unlike Section~\ref{s:drvintsec2}, let us assume that body 2 passes the origin at time $t=0$ with velocity $\V_2$, 
and body 1 passes the origin a time $t=\Delta t$ with velocity $\V_1$.  
(Note that $\Delta t$ here may be negative, if body 1 passes the origin before body 2.)
At some time $t=t_*$, the two bodies achieve a minimum mutual distance.  
For tangential encounters, we can write (without loss of generality) 
\begin{equation}
\V_2 = k\V_1=kv_1(\cos\alpha,\sin\alpha),\qquad -1<k<1. 
\label{e:v1kv2}\end{equation} 
In the vicinity of the origin, we can approximate the motion of body 1 and body 2 as follows,
\begin{eqnarray}
\R_1&=& (t-\Delta t)\V_1 -{1\over2} g(t-\Delta t)^2\hat\x \\
\R_2&=& t\V_2 -{1\over2}gt^2\hat\x = kt\V_1 -{1\over2}gt^2\hat\x.
\label{e:r1r2}
\end{eqnarray}
Then the square of the distance between the two bodies can be expressed as a function of time,
\begin{eqnarray}
|\R_1-\R_2|^2 
    &=& [ (1-k)t -\Delta t ]^2v_{1y}^2  + \{[(1-k)t-\Delta t]v_{1x}+(t-{1\over2}\Delta t)g\Delta t\}^2 \label{e:Dsq}
\end{eqnarray}
The minimum of the mutual distance occurs at $t=t_*$, which can be obtained by the condition $\partial |\R_1-\R_2|^2/\partial t = 0$.  
\begin{eqnarray}
{\partial |\R_1-\R_2|^2\over\partial t} 
&=& 2(1-k)^2v_1^2[1+2\varepsilon\cos\alpha+\varepsilon^2]t \nonumber\\
 & &   - 2(1-k)v_1^2 [1+\half(3-k)\varepsilon\cos\alpha+\half(1-k)\varepsilon^2]\Delta t,
\end{eqnarray}
where $\varepsilon$ is defined as
\begin{equation}
\varepsilon = {g\Delta t\over (1-k)v_1}. \label{e:varepsilon}\\
\end{equation}
Then the minimum distance occurs at
\begin{equation}
t_* = \Delta t {1+\half(3-k)\varepsilon\cos\alpha +\half(1-k)\varepsilon^2 \over(1-k)\den},
\label{e:t*}
\end{equation}
where
\begin{equation}
\den = 1+2\varepsilon\cos\alpha +\varepsilon^2. \label{e:den}
\end{equation}
It is useful to compute
$$(1-k)t_* - \Delta t = -{1+k \over 2}{\Delta t\over\den}(\cos\alpha+\varepsilon)\varepsilon,$$
$$t_* - \half\Delta t = \half {1+k\over 1-k} {\Delta t\over\den}  (1+\varepsilon\cos\alpha).$$
Then, setting $t=t_*$ in Eq.~\ref{e:Dsq}, we find 
\begin{equation}
|\R_1-\R_2|_{min}^2  = (\Delta t)^4 {(1+k)^2g^2\sin^2\alpha \over 4(1-k)^2\den}. 
\end{equation}

For collision to occur, we must have $|\R_1-\R_2|_{min}\leq\tau$, where $\tau$ is the collision radius. 
Thus, collision will occur provided $|\Delta t| \leq \Delta t_{\rm col}$, where $\Delta t_{\rm col}$ is given by
\begin{equation}
(\Delta t_{\rm col})^4 (1+k)^2g^2\sin^2\alpha = 4\tau^2 (1-k)^2 (1+2\varepsilon\cos\alpha+\varepsilon^2).
\label{e:Deltat4}
\end{equation}

Recall that $\varepsilon\propto\Delta t$ (Eq.~\ref{e:varepsilon}), therefore Eq.~\ref{e:Deltat4} presents a 
quartic equation for $\Delta t_{\rm col}$, whose exact analytic solution is possible but tedious.
Provided that $|\varepsilon|\ll1$, we obtain the leading order solution,
\begin{equation}
\Delta t_{\rm col} \simeq \sqrt{ {2(1-k)\tau} \over {(1+k)g \sin\alpha} }.
\label{e:tcol}
\end{equation}
A better approximation can be achieved by plugging the first approximation, Equation~\ref{e:tcol}, into Equation~\ref{e:Deltat4}, to obtain the leading order correction; this yields
\begin{equation}
\Delta t_{\rm col} \simeq \sqrt{ {2(1-k)\tau} \over {(1+k)g \sin\alpha} }  \pm {{\tau} \over {(1+k) v_1 \tan{\alpha}}}.
\label{e:tcol2}
\end{equation}
These two equations above are equivalent to Equation~\ref{tcoltng} and \ref{tcoltng2}.

The condition $|\varepsilon|\ll1$ requires that the change, $g\Delta t_{\rm col}$, in the heliocentric velocity of the bodies over 
the time $\Delta t_{\rm col}$ is much smaller than the encounter velocity, $\Delta v=v_1-v_2$. 
We note that, using this approximate solution in Eq.~\ref{e:varepsilon} we have 
\begin{equation}
\varepsilon \simeq \sqrt{2g\tau\over(1-k^2)v_1^2 \sin\alpha} = \sqrt{\tau/r \over \langle v_T\rangle\Delta v/v_c^2},
\label{epsilon}
\end{equation}
where 
$\langle v_T\rangle = (v_1+v_2) \sin\alpha/2$ is the average transverse velocity of the two bodies 
at the collision point, and
we used $g=GM_\odot/r^2 = v_c^2/r$ ($r$ is the heliocentric distance at the collision point, 
and $v_c = \sqrt{GM_\odot/r}$ is the heliocentric circular velocity).
Thus, the condition $|\varepsilon|\ll1$ is equivalent to  
\begin{eqnarray}
\sqrt{\tau/r \over \Delta v/v_c} \ll 1& \qquad\qquad {\rm (for\,\, prograde)},\\
\sqrt{\tau/r \over 2\langle v_T\rangle/v_c} \ll 1& \qquad\qquad {\rm (for\,\, retrograde)}.
\label{condition}
\end{eqnarray} 
In words, we can state the condition as: the collision radius (as a fraction of the heliocentric distance) is much smaller than the velocity difference $(1-\left| k \right| ) v_1$ (as a fraction of the orbital velocity). In physical terms, we can describe this as the condition that the collision radius $\tau$ should be much smaller than the heliocentric distance, when the velocity difference is not a vanishingly small fraction of the orbital velocity.

\subsection{Derivations for Non-intersecting Orbits}

For the case of non-intersecting orbits, we note that at the location of the MOID, the tangent to each orbit is normal to the relative distance vector, $\s$. We assume that the minimum mutual distance occurs in the vicinity of the MOID location.
Let's choose as origin that point on body 1's orbit where the MOID occurs, and let's choose the plane of body 1 as the XY plane.   

Let's assume that body 2 passes the MOID location, $\s$, at time $t=0$ with velocity $\V_2$
and that body 1 passes the origin at some time later, $t=\Delta t$ with velocity $\V_1$.  
Then, for near-tangential encounters we can write, 
\begin{eqnarray}
\V_i &=&\V_i(\cos\alpha,\sin\alpha,0),\\
\V_2 &=& k\V_1, \qquad -1<k<1.
\end{eqnarray} 

We can also express the time-dependent positions of body 1 and body 2 as follows.
\begin{eqnarray}
\R_1&=& (t-\Delta t)\V_1 -{1\over2} g(t-\Delta t)^2\hat\x \\
\R_2&=& \s + t\V_2 -{1\over2}gt^2\hat\x = \s +  kt\V_1 -{1\over2}gt^2\hat\x , \label{e:nxr1}
\end{eqnarray}
Then the distance between the two bodies is 
\begin{eqnarray}
|\R_1-\R_2|^2 
    &=& s_z^2 + \{[ (1-k)t -\Delta t ]v_{1y} - s_y \}^2  \nonumber\\
     & &   + \{[(1-k)t-\Delta t]v_{1x}+(t-{1\over2}\Delta t)g\Delta t - s_x \}^2 \nonumber\\
    &=& [(1-k)t-\Delta t]^2v_1^2 +(t-\half\Delta t)^2(g\Delta t)^2 \nonumber\\
    &  & + 2(t-\half\Delta t)[(1-k)t-\Delta t]v_{1x}g\Delta t 
	- 2(t-\half\Delta t)s_xg\Delta t +s^2
\label{e:nxDsq}
\end{eqnarray}
The minimum of the mutual distance occurs at $t=t_*$, which can be obtained by the condition $\partial |\R_1-\R_2|^2/\partial t = 0$.  
First, we derive
\begin{eqnarray}
{\partial \over \partial t}|\R_2-\R_1|^2 
    &=& 2[(1-k)t-\Delta t](1-k)v_1^2 +2(t-\half\Delta t)(g\Delta t)^2  \nonumber\\
    & & +2[(1-k)t-\Delta t+(1-k)(t-\half\Delta t)]v_{1x}g\Delta t - 2s_xg\Delta t \\
    &=& 2(1-k)^2v_1^2(1+2\varepsilon\cos\alpha+\varepsilon^2) t \nonumber\\
     & &   -2(1-k) v_1^2[1+\half(3-k)\varepsilon\cos\alpha +\half(1-k)\varepsilon^2+{gs_x\over(1-k)v_1^2}] \Delta t
 \label{e:nxpDsq}
 \end{eqnarray}
where $\varepsilon$ is given by Eq.~\ref{e:varepsilon}.
Then, we find 
\begin{equation}
t_* = {\Delta t\over(1-k)\den} [1+\half(3-k)\varepsilon\cos\alpha +\half(1-k)\varepsilon^2 +{gs_x\over(1-k)v_1^2}],
\label{e:tstar}
\end{equation}
It is useful to compute
$$(1-k)t_* - \Delta t = -{1+k \over 2}{\Delta t\over\den}[\varepsilon\cos\alpha+\varepsilon^2-{2gs_x\over(1-k^2)v_1^2} ],$$
$$t_* - \half\Delta t = \half {1+k\over 1-k} {\Delta t\over\den}  [1+\varepsilon\cos\alpha +{2gs_x\over(1-k^2)v_1^2} ].$$
We use Eq.~\ref{e:tstar} in Eq.~\ref{e:nxDsq} to find that the minimum distance: 
\begin{eqnarray}
|\R_1-\R_2|_{min}^2 
&=& s^2 
+{(1+k)^2(\Delta t)^2v_1^2\over4\den^2}\Big\{
      (\varepsilon\cos\alpha+\varepsilon^2-2\lambda)^2 \nonumber \\
& & +(1+\varepsilon\cos\alpha+2\lambda)^2\varepsilon^2
    -2(1+\varepsilon\cos(\alpha) + 2\lambda)(\varepsilon\cos\alpha+\varepsilon^2-2\lambda)\varepsilon\cos\alpha \nonumber\\
& & -4\lambda(1+\varepsilon\cos\alpha+2\lambda)(1+2\varepsilon\cos\alpha+\varepsilon^2)\Big\},
\label{e:nxDminsq}
\end{eqnarray}
where 
\begin{equation}
\lambda = {gs_x\over(1-k^2)v_1^2}
\label{e:nxdelta}
\end{equation}
In a similar way to Equation~\ref{epsilon}, the small parameter $\lambda$ can be understood as
\begin{equation}
\lambda = {{s_x /r} \over {2 \langle v \rangle \Delta v / {v_c}^2}},
\end{equation}
where $\langle v \rangle = (v_1+v_2) /2$ is the average velocity of the two bodies. Because $s_x \le \tau$, we see that $\lambda$ is of order $\varepsilon^2$.

Setting $|\R_1-\R_2|_{min}^2 =\tau^2$, we obtain a polynomial equation for $\Delta t_{\rm col}$:
\begin{eqnarray}
& & {1 \over 4}{(1+k)^2(\Delta t)^2v_1^2}\Big\{
      (\varepsilon\cos\alpha+\varepsilon^2-2\lambda)^2 
    +(1+\varepsilon\cos\alpha+2\lambda)^2\varepsilon^2 \nonumber\\
& &   -2(1+\varepsilon\cos(\alpha) + 2\lambda)(\varepsilon\cos\alpha+\varepsilon^2-2\lambda)\varepsilon\cos\alpha 
    -4\lambda(1+\varepsilon\cos\alpha+2\lambda)(1+2\varepsilon\cos\alpha+\varepsilon^2)\Big\} \nonumber\\
& &\qquad\qquad\qquad\qquad = (\tau^2-s^2)(1+2\varepsilon\cos\alpha+\varepsilon^2)^2
\end{eqnarray}
This is a polynomial of the 6th degree in $\Delta t$.  Keeping only the leading order terms, we need only solve a quadratic in $(\Delta t)^2$:
\begin{equation}
{1\over4}{(1+k)^2g^2\sin^2\alpha\over(1-k)^2}(\Delta t)^4
-{(1+k)gs_x\over 1-k}(\Delta t)^2
-(\tau^2-s^2) = 0
\end{equation}
Only one of the two solutions is physical; we find
\begin{equation}
\Delta t_{\rm col} = \sqrt{2 \tau (1-k) \over g \sin{\alpha} (1+k)}  \left( \sqrt{1 -  \frac{s^2}{\tau^2} \sin^2{\beta}}  - \frac{s}{\tau} \cos{\beta}  \right)^{1/2},
\label{e:nxDtsq}
\end{equation}
which is the same as Equation~\ref{DtTng2}.

\end{document}